# False Asymptotic Instability Behavior at Iterated Functions with Lyapunov Stability in Nonlinear Time Series

Charles Roberto Telles[1]

[1] Director Board, Research Advisory. Secretary of State for Education and Sport, Curitiba PR 80240-900, Brazil
charlestelles@seed.pr.gov.br

**Abstract.** Empirically defining some constant probabilistic orbits of $f(x)$ and $g(x)$ iterated high-order functions, the stability of these functions in possible entangled interaction dynamics of the environment through its orbit's connectivity (open sets) provides the formation of an exponential dynamic fixed point $b = S_{n+1}$ as a metric space (topological property) between both iterated functions for short time lengths. However, the presence of a dynamic fixed point $f(g(x))_{x+1}$ can identify a convergence at iterations for larger time lengths of $b$ (asymptotic stability in Lyapunov sense). Qualitative (QDE) results show that the average distance between the discontinuous function $g(x)$ to the fixed point of the continuous function $f(x)$ (for all possible solutions), might express fluctuations of $g(x)$ on time lengths (instability effect). This feature can reveal the false empirical asymptotic instability behavior between the given domains *f* and *g* due to time lengths observation and empirical constraints within a well-defined Lyapunov stability.

## 1 Introduction

The first main problem addressed in this research is to establish a difference between empirical experiments and theoretical mathematical simulations regarding asymptotic stability and bifurcation phenomena [1]. Although asymptotic iterated functions can be often considered empirically invisible in their nature of proportionality and convergence [2,3,4] due to time dependent outcomes and stability scenarios where high level of complexity is present in a given phenomenon and as noted by Newman *et al* [4], all the solutions in convergence to a dynamic and time-varying fixed point can be empirically visible if time nonlinear lengths of phenomenon can be accelerated by mathematical tools or physical properties in connectivity with the event [3].

Based on this methodologies views, a given phenomenon and analysis always can point to the correct observation of an event as far as mathematical frameworks are used to fit in the event. However, this framework can be unsuited to empirical analyses where perturbations over the initial conditions and during phenomenon expression are possible to occur or asymptotic effects have its occurrence not defining a specific probabilistic distribution over time [4,5,6]. Following this path, the opposite aspect of the mathematical and theoretical framework can be given, when in one empirical analysis, the data is analyzed and a mathematical model is fitted to it, leading to a conclusion where events occur as something independent [3] of the empirical dimension of analysis. In this view, theoretical schemes that don't show any different aspect other than the initial formulation and mathematical prediction of a given phenomenon can't be a good model for nonautonomous equations, where empirical constraints can be the cause of bifurcation phenomena [1,4,7].

Differing theoretical qualitative partial differential equations analysis and empirical predictions or experiments towards event's convergence and stability behaviors, the frequency of iterations with which these functions occur partially on time intervals in their theoretical formulations (as time invariants and autonomous) or empirical observations (nonautonomous and time-varying), affecting the physical nature of a phenomenon, are subject to observation and discrimination to the extent that they can be empirically observed, knowing the metric of spaces that constitute all the stages of a given event in terms of its iterated functions be a product of complex interactions [2,3,7,8]. Concerning qualitative analysis of partial differential functions, mathematical theoretic parameters might be defined without the observation of physical and time constraints, whose properties of events in evolve, only in the light of an empirical experimentation, could express unpredicted mathematical expectation, and thus differing from empirical data expressions. However, as the metric space of these iterated functions oscillates in their expressions considering an empirical view of phenomena and can increase distances from each other in terms of possible complex interactions, it is possible, deductively, to describe the event as presenting a dynamic asymptotic instability between functions as a constant defined as $b$ in terms of its expression at short time lengths for all finite solutions directed to a given fixed point as noted by Lundberg (1963) [9]. But considering the same phenomenon as defined in this research, as a dynamic exponential fixed point, that evolves as far as instability turns it into new patterns of stability, resembling Williamson's concept (1991) [4,10], the phenomenon in larger time lengths, inductively [3], might present stability evolution (convergence) in the Lyapunov sense of stability [11]. In this sense, for identifying at Lyapunov stability an asymptotic stable behavior, for empirical observations, it is necessary to describe an event considering its escaping and approximation orbits similarly to a KPZ Brownian aspect of convergence [12] or the stochastic Lorentz system [13] in order to make visible, mathematically and empirically, the asymptotic stability formation feature within the apparent unstable dynamics. However, in this research it will be demonstrated that the apparent asymptotic instability observed in nature's function expression in connectivity and time lengths (iterated and created by high level of variable's possible metric spaces within disjoint open sets) is in true a false asymptotic instability effect. This feature occurs as far as the time interval of phenomenon observations confers to a given fixed point, oriented as an exponential function of $f(x)$ and its complex not *i.i.d.* string of variable's interactions $g(x)$, a quantized empirical result that will express oscillatory overall proportionality (not the asymptotic equipartition property) and the overall convergence for all dynamical fixed point observed at larger time lengths within a nonlinear time series.

The camouflage effect of instability resides in the sense that the asymptotic instability definition in terms of empirical observation does not match with true functions correlation if compared with the theoretical view of mathematical simulations, being the effect of time lengths a tool in which empirical results can present massive quantified results of a function that can be asymptotically stable as the whole system observations or in an exceptional case, even be not asymptotic at all.

It is defined as a dynamic fixed point in this research the formation by iteration and interaction between distinct iterated functions, on a metric space in which one of the function $g(x)$ that are defined partially by a fixed point $f(x)$ assume other distances between two points of the two considered functions, alternating the position in the complete existent metric space between one function $f(x)$ and the other subsequent iteration of $f(g(x))$ as a nonempty space that represents in empirical terms the connectivity of $f(x)$ and $g(x)$ as an evolution and topological expansion of the system defined as the constant $b$, thus generating randomly and dynamical exponential fixed points as the $b$ constant.

This connective metric space in turn generates a kind of instability between the fixed point metric position that can be defined in a specific order of possible orbits of interactions in the system for $f(g(x))$ (solutions proximity) [14] and at the same time a region of space in which the iteration reaches their maximum degree of expression presenting higher distance between two points of overall iterated functions (escaping solutions) [15-18].

Thus, Banach's fixed-point theorem objectively illustrates this instability distribution of iterated functions over a single fixed point in a complete metric space, considering for it the dynamic fixed point concept and reflecting this definition as one of the iterated functions partial attraction to a fixed point for each interaction or for some time lengths of event, thus creating a connectivity metric space [12,13], however generating expansively many distinct fixed points.

This research practical implications are that, empirically, many results presented by scientific community show that a given studied system present asymptotic instability, due to the limitation of time acceleration of events. This aspect can, empirically predicts, how physical phenomena will behave in very complex scenario for many fields of science and also for policy making regarding descriptive statistics methodological limitations [19,20] in the big cities administration. The main result of this research dealt with a simple example using a nonlinear time series data, where an approaches for nonautonomous techniques were done, giving a general overview of time-varying solutions convergence at nonautonomous partial differential equations.

## 2   Methodology

### 2.1   Connectivity and iterated functions.

**Theorem 1.** Consider within a nonautonomous system of partial differential equations, empirically designed, of one iterated and continuous function *f* defined as a mirror (fixed point) of the discontinuous iterated function $g$ and having as its derivative $g$ constantly iterated in connectivity to $f$ as a constant $b$. A complete metric space as $\delta = (x_{n+1})f(g(x))$ as the constant $a$, present all solutions in convergence as $\varepsilon = y = b$, but presenting high oscillations as time expands the event, hence expressing asymptotic instable behavior caused by $a$ derivative and the connectivity topological property. These two iterated functions $f(x)\ and\ g(x)$ in terms of their physical nature have defined, but high complex probabilistic behavior, and only in the time acceleration of the phenomenon it is possible to observe that the distributive instability asymptote behavior is rather an effect of unobserved overall $g(x)$ convergence towards $f(x)$ or in other words as a dynamical fixed point as the constant $b$ with overall possible convergence of the constant $a$.

In this sense, consider for the function $f(x)$ the sum of all possible metric spaces produced by the complex interactions of variables present at $g(x)$, and thus defining itself as the fixed point of convergence, as Picard–Lindelöf theorem, like $\varphi_0(t) = f(g(x))_0$, hence defining the $a$ constant of convergence for $g(x)$. As far as $f(x)$ define itself as a product of $f(g(x))_{x+1}$, as $\varphi_{x+1}(t) = f(g(x))_0 + \int_0^{x+1} g(x)(s, \varphi_x(s))ds$, being $s$ a local uniqueness for the iterative behavior, a fixed point is created for each string of iterated events and it has as its derivatives the partial differential equations of $g(x)$, where this later function assumes a dynamical time-varying probabilistic variance that can be observed in terms of exponential empirical behavior caused by complex interactions among $g(x)$ multivariable functions (not embedding). The $a$ constant of convergence, by $g(x)$ expansion, leads $f(g(x))_{x+1}$ to assume an empty space within the disjoint sets, thus generating a connectivity metric space and topological property defined as the $b$ constant, or in other words $f(x)_x$ like time-varying dynamical fixed points. In this sense, fixed points can be defined for all $f(x)$ constantly, defined as a Banach fixed points, and for $g(x)$ all solutions goes near $f(g(x))_{x+1}$ dynamical fixed points, but can remains partially tangent to $f(x)_x$ formed fixed points for all time, constantly as $b$ undefined asymptotic behavior (theoretically speaking). Note that the notation $f(x)_x$ is equal $f(g(x))_{x+1}$, except for the iterative aspect. However, in this research the notation $f(x)_x$, will be replaced by $f(g(x))$ for the next explanations.

Note that this trivial expression of $a$, in the light of an empirical explanation, is the desirable solution of the event, being this expression, the most observed solution of a given phenomenon, however, not observed as an $b$ constant due to physical and experimental constraints that will lead the observation of the nonautonomous partial differential equations properties.

The time of occurrence $x(t_{n+1}) \geq 0$ for each iterated event representing each function in connectivity ($b$) is constant at an expression frequency of time as $t_{n+1} + t_{n+1}$ and probabilistic distribution as a solution $\varphi = P_n$, with trajectories $X(t_n)$ as input data, the solutions $\varphi(t_n)$ assumes asymptotic stability of $X(t_n) - \varphi(t_n) < \delta$ for each single iterated event, thus reflecting the formation of a constant already identified as $a$. The interaction of these functions in a given region of the metric space $x_{n+1} = f(g(x))$ generates, assuming that the quantitative properties of the event remain with partial and asymmetric numerical transformations to their original form ($\partial$), for each interaction such as,

$$\frac{\partial x_{n+1}}{\partial g}(f,g) = or \leq 1 \therefore x_{n+1}(f,g) = f(g(x)) \tag{1}$$

a solution to a fixed point at $y$ axis that defines itself as a constant and random variable $\partial x_{n+1} = b\{S_1, \dots, S_n\}$ (dynamic fixed point) for the function $f(g(x))$ for each one of the iterations. The distance between two points in each iterated function at $y$ as $f(g(S_n))$ occurs in a general and the maximum empirical expansion as,

$$d(f(x), g(x)) = d((g(x) = fx + gx) \leq f(x) \text{ or } \leq \sum_{S_n+1}^{n} S_n \tag{2}$$

for short time intervals (embedding properties). Thus confirming the Lyapunov sense of asymptotic stability [11] defined as expressing a high convergence rate to the defined fixed point. Visualizing that scenario as time passes, the function's behavior does not change under small input perturbations the asymptotic stability of phenomena at each of the iterations, where $x_{n+1}$ can be represented by a connectivity in which the global distance between two points of function $f(x)$ and $g(x)$ domains assume distances equal to,

$$d(f(g(S_n)), g(x), f(x)) \leq d\left(\sum_{S_n+1}^{n} S_n, f(g(S_n))\right), \qquad (3)$$

being the global distance presenting high vector instability and therefore as the overall distance of the constant $a$ start to increase, the dynamical fixed point for $f(g(x))$ start its formation, thus generating the already defined constant $b$. The overall convergence of the function $g(x)$ when attracted to a fixed point of $f(x)$ can be defined, roughly, as $\lim_{n \to \infty} f(x) = \sum_{S_n+1}^{n} S_n - g(x) = 0$ such that $x(S_n + 1) < \delta$, for each iteration $S_n + 1$, when considering only one fixed point condition and $S_n + 1$ is a metric space in which $g(x)$ is not defined at the fixed point mutually with another iteration as indicated in number notation (4).

$$\begin{aligned} f: f(g(S_n)) &\to S_{n+1} \\ g: g(f(S_n)) &\to S_{n-1}. \end{aligned} \qquad (4)$$

In the opposite direction of the $f(g(x))$ pattern formation, the same function as now the constant $a$, the dynamic fixed point, can be defined where compared to the previous given equation, would necessarily assume the monotone definition as $\lim_{n \to \infty} f(x) = \sum_{S_n+1}^{n} S_n - g(x) > 0$, hence defining itself mutually among large time intervals as $x(S_n + 1) < \delta$, being $\delta$ redefined constantly as far as $f(g(x))$ assumes new states of oscillation and convergence, not assuming a behavior as described in (4).

And also $g(x)$ by its turn has its metric space among complex variables interactions defined by nonlinear dynamics of event, not being possible to measure how the system input and output evolves for larger time lengths as a nonautonomous functions. Despite of a constant iterated convergence ($a$), $g(x)$ assumes $\mathbb{R}$ solutions that empirically expand into nonlinear time series, thus not presenting any visible pattern of oscillation at larger time intervals. This feature resembles the nonlinear time series as pointed by Kantz and Schreiber [21] of the partial differential equation's system.

This complex and random property of variable's interaction of $g(x)$ means that to the extent that $f(x)$ is dynamically defined by fixed points for each iteration, there is the formation of fixed positions in time as $f(g(x))_{x+1}$, with which the nonlinearity of $a$ distance between two points of both sets for each iteration assumes a Lyapunov sense of asymptotic instability [11,22] with a bounded and embedded mapping condition where for all vectors, the solutions will always be less than or equal to the sum of all dynamic fixed points ($v \leq \sum_{S_n+1}^{n} S_n$), therefore presenting pattern formation for shorten time lengths. When observing the connectivity $b$ of the iterated event, this characteristic reveals that the maximum dynamical fixed points are formed as an exponential function of $b$ raised by $a$ exponent of the iterated functions $f(x)$ at $y$ axis and $g(x)$ vectors functional expression at $x$ axis, generating a defined phase space, locally stable for every $y$ point and unstable considering all $y$ nonlinear time series.

According to the fixed point position of the phenomenon generated by the string sequences (orbits) of interaction between complex variables of $g(x)$ and the average distances generated between the fixed points convergence for each iteration, this behavior of the event also allows us to observe, as will be described in the results section, that the distance for all solutions in all iterated events of the nonlinear time series in the dynamics of maximum and minimum metric spaces, assume an asymptotic instability that accompanies certain empirical expressions that is the not observable stability convergence effect [22], in which strings of iterated functions in connectivity have with the previous and posterior string events. This feature may not be visible (discretized) in terms of overall convergence or non-convergence due to an asymptotic camouflage effect.

This phenomenon can best be described as an orderly sequence of iterations through time as $x_{n+1} + t_{n+1} = 1$ with parameters determined in $a$ as $S_n + 1$, $P_n$ and $d$ in the relationship that is defined only from $b = S_1, \ldots, S_n$, generating a composition like $(f \circ g)(S_n)$ as $f(S_n + 1) = \{b(S_n) | S_n \in S_n + 1\}$.

A linear prediction could be obtained if for certain strings of $a$ and $b$, defined as Markovians [17], the functions $S_n + 1$ be infinitely discretized by presenting growth or decreasing non exponential value caused by $a$ in the available proportion of $S_n + 1$, revealing the flow of process iterations as $\int_{g(x)}^{f(x)} (f^m \circ g^n) = (S_n^{m+n})$ as possibly ordinary differential equations in its various position for every single fixed point and for the formation of other fixed points. This circumstance in the opposite objective of this research, can be defined as a linear time series where statistically it is possible to determine the direction of all vectors in the field [19,23].

To visualize the interaction of complex variables in iterated functions and the formation of asymptotic instability, the following sets are defined as a stylized example in figure 1.

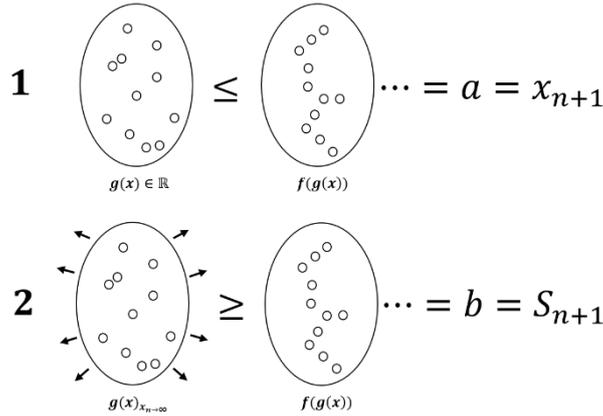

**Fig. 1.** Formation of dynamic fixed points by connectivity $(f \circ g)(x)$.

**Lemma 1.** It is possible to observe that the fixed points $S_n$ are generated from the $S_n + 1$ interactions $g(x)$ and iterations $f(x)$ event. Thus, event discretization $S_n +$ 1 occurs as a string sequence (trajectory) created from the availability of $S_n$ of being constant, dependent of $g(x)$ evolution, but randomly formed from $g(x)$, which these fixed points in turn assume a function defined as $(f \circ g)(x)$. Therefore, for both functions $f$ and $g$, the iterations of each function remain as an image of interactions occurring at a given moment from $n$ fixed points $S_1, \ldots, S_n$ randomly or not, generated as $f:(S_n + 1) + (S_n + 1) \to S_n \therefore f \times g$, where, considering the larger time lengths, the metric space of connectivity of the constant $b$ between two points of each iterated function is defined as $d(f(g(S_n)), g(x), f(x)) \leq or \geq d\left(\sum_{S_n+1}^{n} S_n, f(g(S_n))\right)$, expansively compared to the definition of (3), the global distance, where it was considered short time lengths for the iterations as observed analogously by Ignat'ev *et al* [6]. If this definition be considered in the view of a common linear time series, the whole distance would be defined as (3), even for larger time lengths, due to the existence of only one fixed point as well as average distance and convergence rate among each iteration.

**Proof of theorem 1 and lemma 1.** The equation that describes the behavior of figure 1 for each iterated function $f(x)$ in terms of time can be described as:
$$b = f(x) = S_n + 1 = (x_{n+1} + t_{n+1})P_n \tag{5}$$
Since all variables of each function have defined values where $x \in \mathbb{R}$, then the interaction orbits of $g(x)$ have heterogeneous variables (discrete and continuous) interactions, thus presenting as a limit to $f(g(x))$ as a maximum metric space of heterogeneity. These characteristics of $f(g(x))$ as a dynamical fixed point formed as far as $g(x)$ converges to $f(x)$ for each iteration and as a solution of $g(x)$ specific pattern random formation (discrete and continuous variables), the string sequence of iterations can be defined as found in Lyapunov's stability, constantly changing for a new $g(x)$ towards $f(x)$, like,
$$(S_n + 1)n = (x_{n+1} + t_{n+1})P_n, \text{ as the constant } b \tag{6}$$
for each $f(g(x))$ originated, where, for each $g(x)$, the solutions at each iteration presents,
$$g(x)(t_n) - \varphi(t_n) \leq \delta, \tag{7}$$
as a nonautonomous PDE (partial differential equations), or in other words, an adaptive and self-organizing system.

At this point, it is mandatory to note that as briefly described in the introduction section, the mathematical theoretical observation of phenomena can't induce by its nature of analysis, that the Lyapunov stability is rather fully unstable in the view of empirical results. Also, this feature, in the view of time acceleration of event, it project an awkward empirical observation of an unstable convergence of $g(x)$ towards $f(x)$ (not confirmed at the mathematical point of view) and totally no pattern formation as observation collect data sets of phenomena. Exemplifying it, this feature can be empirically viewed when specimen evolution is theoretically observed through the Darwinian framework of analysis [24].

The empirical observation of asymptotic instability, now better can be defined as the empirical observation of non-convergence of a given phenomenon, can be filtered in a nonlinear time series investigation by time acceleration of the event in to occur. This mathematical or empirical possibility could lead to a new iterated event's projection defined as the sum of all dynamical fixed point, that are being observed as unstable in most of the nonlinear time series data. The time acceleration of event would lead to the function defined as a constant $a$ where exponentially defines by its turn the constant $b$ like the expressions,

$$a = f(x) = \sum_{S_n+1}^{n} S_n - g(x) \text{ or } \sum_{S_n+1}^{n} S_n = f(x) + g(x) \text{ or } g(x) = -f(x) + \sum_{S_n+1}^{n} S_n$$
$$b = \sum_{t_n+1}^{n} S_n = S_1 + S_2 + S_3 + \cdots + S_n \geq 0. \quad (8)$$

Where for all $g(x)$ at each $S_n$, there will be necessarily a value for $g(x)$ that is not equal 0 hence predicting the system complexity for each iteration as in convergence to $f(x)$ (asymptotic stability) for each iteration and at the same time the connectivity of iterations reveals in the light of empirical observation a growing or decreasing effect towards $f(x)$, but overall asymptote stability towards $(x_{n+1} + t_{n+1})P_n \to \infty \therefore f(x) \sim g(x)$.

For the third equation (8) form $g(x) = -f(x) + \sum_{S_n+1}^{n} S_n$, the empirical nonlinear time series express internal movements of which unpredictable trajectories are observed constantly as far as the system is perturbed by strong initial input of new variables within the system (system expansion).

Considering the constant $b$, roughly, as $g(x) = -f(x) + \sum_{S_n+1}^{n} S_n$ as unpredictable, but in convergence to the $f(g(x))$ fluctuations, for the constant $b$ for only one iteration at a time, the empirical observation of phenomenon for the asymptotic stability effect can only be observed if considering $g(x)$ as derivative of $f(x)$ where this later function is assumed to be non-exponential in the sense of Lyapunov convergence, hence without nonlinearity aspects.

In another sense, for all derivatives $g(x)$ that are produced for $f(x)$ as a result of strong asymptotic instability and not only for a single observation or short time lengths observations, the iteration process might be better explained in its oscillations through an exponential function defined for $f(g(x))$, where the convergence rate of distant metric spaces between $f(x)$ and $g(x)$, created in nonlinear time series observation, can be obtained as the overall growth or decrease of each iteration and nonlinear solutions produced by $g(x)$ as,

$$\left(f(g(x))_1, f(x)_1\right), \left(f(g(x))_2, f(x)_2\right), \dots, \left(f(g(x))_n, f(x)_n\right), \quad (9)$$

in the form of,

$$-\sum_{S_{n+1}}^{n} f(g(x))_{x+1} e^{bS_n} + a \sum_{S_{n+1}}^{n} e^{bS_n} \quad (10)$$

as the $b$ constant and exponential form, and,

$$\sum_{S_{n+1}}^{n} f(g(x))_{x+1} S_{n+1} e^{bS_n} - a \sum_{S_{n+1}}^{n} S_n e^{bS_n} \quad (11)$$

as the constant $a$ with undefined exponential rate for $b$, due to complex variable's nature (discrete and continuous).

Q.E.D. □

## 3 Results

### 3.1 Dynamical fixed points and connectivity metrics.

Following this proof, all given possible solutions directly reflect the maximum number of locally stable and unstable maps [22], being it the trajectories of the constants $a$ and $b$ generated by the functions $f(x)$ and $g(x)$ on time. However, it should be noted that in the expression $S_n + 1 = f(x) + g(x)$ or $f(x) - g(x)$, it can be converted into $\int_{g(x)}^{f(x)}(f^m \circ g^n)$, where the sum of existing function $g(x)$ can reach as many iterations as possible in the unbounded event, and many of the iterations can be identical or not. It is not possible to observe these event's variations in an empirical sense as far as the time length considered for analysis is not enough to express asymptotic exponential stability. Thus, although the asymptotic exponential stability is not visible at short time lengths, both functions $f(x)$ and $g(x)$ clearly differ in the initial condition of the event in terms of empirical properties of the iteration and its expressions within a context, for example, probabilistic distribution (figure 2) for linear / nonlinear events which directly affects the ability of the observer to discretize the event and classify it accordingly. So when in $f(x) \sim g(x)$ in terms of solutions, it is obtained a greater metric space among iterations and consequently $g(x)$ assumes a greater distance from the fixed point of $f(x)$ when considering the initial conditions of the phenomena. And in the opposite hand, if $g(x) \sim f(x)$, the uniqueness of the system can be observed in short time lengths, being this feature observable empirically at the beginning of experimentations, but not necessarily mean a deterministic conclusion.

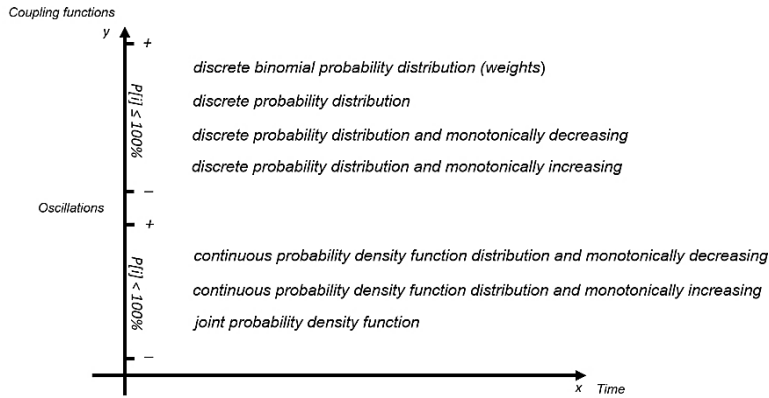

**Fig. 2.** Probabilistic distributions occurring in the complex variables due to coupling effects, where $P[i]$ represents the deterministic to stochastic strings of distribution for a given event $i$ to happen with 100% or less probabilities.

These statements are very important for determining empirical phenomenon behavior and also many researches can possibly present failures in their findings since no time acceleration is possible to obtain and disprove asymptotic instability observations as it can be seen as one example in the figure 3. Figure 3 represents the consumption of water by public schools during the period of 2007 until Oct., 2019, month by month [25]. During these nonlinear time lengths, it is possible to observe the oscillations of $a$ constant that redefines $b$ during the years. The exponential function (horizontal line) of $b$ constant presents a gradual water consumption growth that is not observable and not correlated if considering previous year's data or short time periods data as an average or most frequentist data. In this phenomena characterization, the asymptotic instability observed during the years, thus express a stability feature that follows all dynamic fixed point solutions at each pattern formation of convergence and stability. The relative convergence of overall variables towards $b$ constant reflects also the system adaptation to internal and external variables that influences the system modifying the $a$ constant with deep instability.

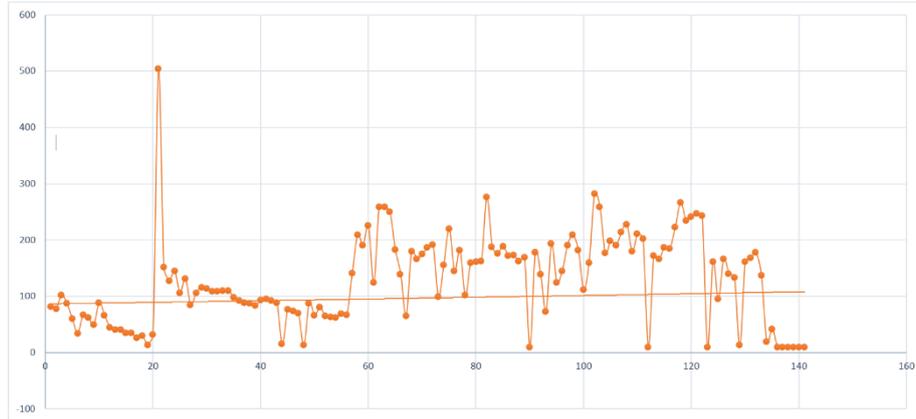

**Fig. 3.** Expression frequency of interactions and iterations on time between iterated functions from a dynamic fixed point in connectivity.

If the frequency with which specific fixed point attracts an iterated function is set to be equals for all $S_n$ for a long period (considering time here as relative to the phenomena observed), hence exhibiting the same behavior as described in the methodology section as a Cartesian interaction product (linearity of functions behavior), the false asymptotic instability effect ceases as the distances between dynamic fixed points for each of the interactions in connectivity are no matter larger or short. Some resource usage, depending on the context and variables that generate the event, have this function performance, not equal to figure 3 properties.

One of the most difficult aspects of determining a fixed point convergence within nonlinear dynamics time series, would be if both functions and iteration frequencies towards instability do not match in either position $S_n$, $S_{n+1}$ and $S_{n-1}$ for any kind of connectivity aspect in $f(g(S_n))$ (attraction) or $g(f(S_n))$ (pattern formation).

The main point here is to make the difference between theoretical and empirical observations. For the first, a mathematical framework can lead to the correct observation of a given phenomenon as far as defined initial conditions are existent. But consider now an empirical sense of observation. The frequency that it occurs on time for input and output within a control system, where for large periods of time the event does not change as well as its initial conditions leading it to the notion of Lyapunov stability, but as for the input variables, if the event present high frequency and nonlinear behavior of oscillations, it can assume an exponential behavior, thus differing from the perfect mathematical framework with only theoretical analysis. The figure 4 displays all variables involved in water usage in public schools within a sample of 2143 distributed in a geographic region of 199,315 km². Just as an example, this figure illustrates the questions of how to define factors of iteration, interaction, frequency, time, space and other physical property aspects of each one of the 35 variables and each possible composition of variables. Considering that each unit has a specific time-invariant dynamic and there is no formation of determined patterns for the variables as an initial condition, therefore, thinking in public administration where big cities will be the challenge to support massive complex interactions, it is possible to scope the issues of how to define a policy making for all administrative units based in a pattern formation that is ranged for all samples with undefined pattern formation or at least if it is possible to make a classification of groups of units that share same pattern behavior; what is the ideal sequence, how to control, which dynamics assumes, what variance, which variable has the greatest influence on each other and how often does it occur.

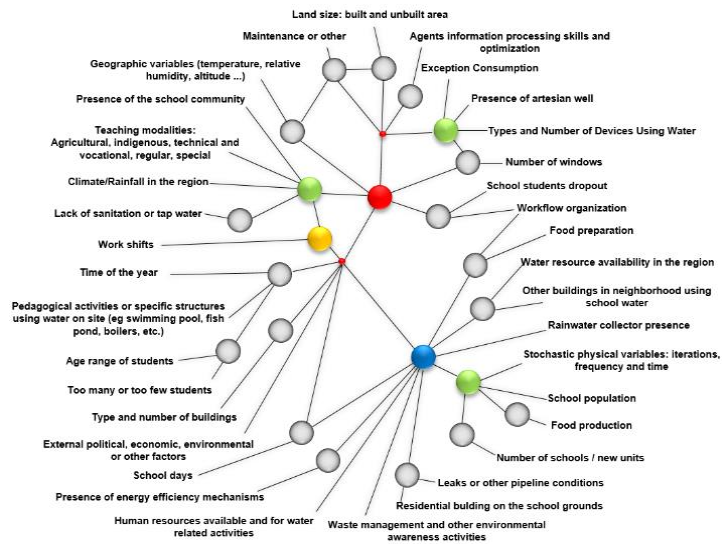

**Fig. 4.** Several configurations might appear with distinct strengths at each variable connectivity and coupling effects. What assumes a given probability distribution to one sample, in another one can assume complete different type of distribution, hence modifying phenomena functioning and composition.

The behavior of variables of figure 4 can also be illustrated by the figure 5, where $g(x)$ can produce strings of variables defined heterogeneously within different time intervals, possibly generating nonlinear pattern formation through time lengths or at least, as this research an asymptotic stability effect towards the whole system as an objective pattern of complex configuration. Note that this figure was used to illustrate the observation of the phenomena with graphical aids.

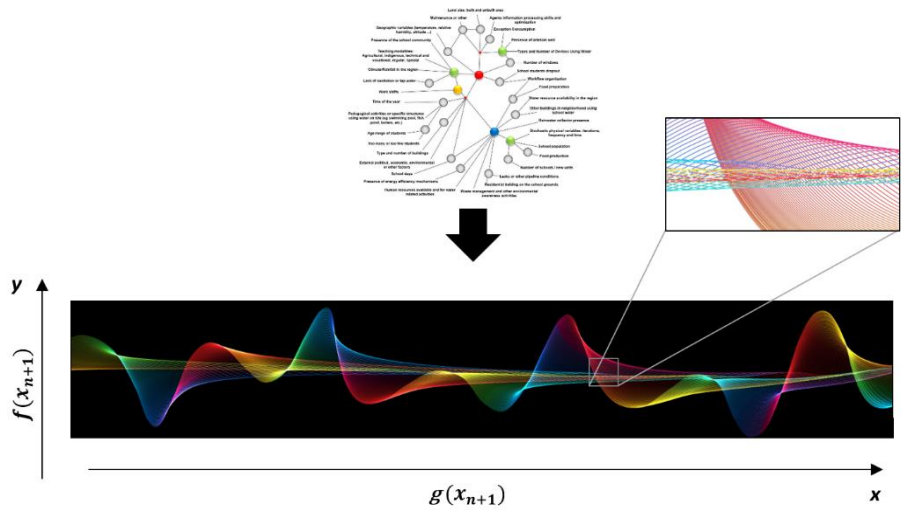

**Fig. 5.** Oscillatory strings of complex variables and the coupling functions of $f(x_{n+1})$ and $g(x_{n+1})$, represented by a stylized graph. In this ideal representation, for practical observation purposes, the strings for each $g(x)$ iteration are assuming relatively harmonic and symmetric oscillations through time intervals, hence reflecting this behavior towards $f(x)$ relative stability and $f(g(x))$ fixed point convergence. The constant $b$, for visual reference, can be observed as a continuous string through oscillations, being composed by complex variables of $g(x)$ and structured by $f(g(x))_{x+1}$ dynamical fixed point convergence.

The scenario represented by figure 5., in the opposed direction can be viewed in figure 6 as representing asymptotic instability, therefore, reflecting the figure 3 phenomena and figure 4 variables, as an example of how the strings of variables under time lengths might assume oscillatory behavior being complicated to generate the same behavior as defined in figure 5.

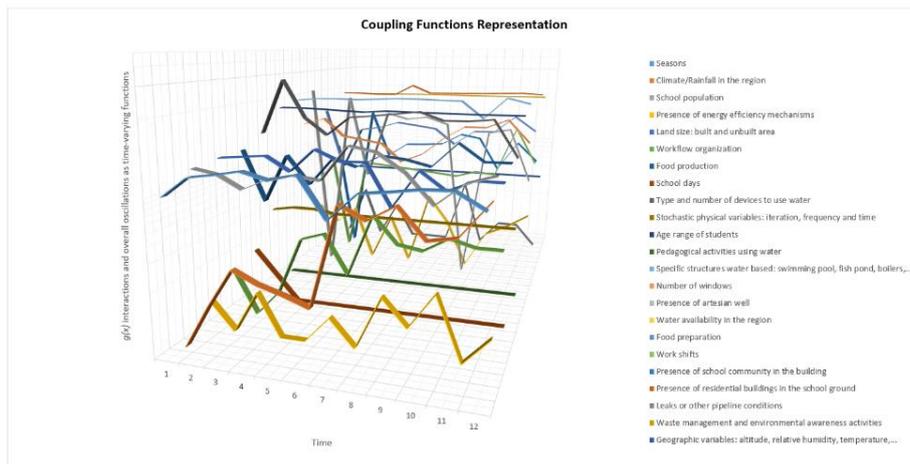

**Fig. 6.** Representation of a strings of variables (as an example) that assume time-varying probability distributions and several possible synchronization states. In order to adjust

the nonlinear system, it is necessary to understand a possible pattern formation of the variable's oscillations and promote modifications at the macroscopic dimension of the system, being it, depending on the considered system of analysis, composed of organic and inorganic components.

Note that this same results approximation can be seen at Monte Carlo calculations, but differing in terms of finding how the mathematical expectation of the not *i.i.d.* complex variables iterations and interactions of $g(x)$ might express as outcomes towards time and system own coupling functions dynamics [26], and thus not defining any visible weight to the probabilistic distributions oscillations (visible by a central limit theorem). This approach is, for example, different from the Tang [27] framework, that identifies weights and the correlation of the system with the central limit theorem, thus resembling the uniqueness concept of stability.

Also, this interactive system and probabilistic distributions share not only physical components within the coupling relations, but there are biological agents [26,28] that causes in many distinct aspects, influences over system functioning. In the view of modern scientific breakthrough, analyzing nonlinearity in the light of public administrative policy and infrastructure [29] are demands of investigations that can constitute a path to establish complex solutions for social systems that present nowadays high level of non-convergence and artificial oscillatory dynamics.

### 3.2 False Asymptotic Instability at Iterated Functions

Despite of a false asymptotic instability effect might exist in an iterated function analysis, this research points out that since time lengths poses the true camouflage effect of the observed phenomena, it can be also used as a tool to design new results towards the quantitative aspect of physical, chemical or biological properties of reality [4,8].

Considering all the nonlinear time series given in the figure 3, the average largest and smallest distances of iterated functions from the dynamic fixed point in $f(x)$ can be obtained by viewing the maximum distances between $f(x)$ and $g(x)$ by the exponential function defined as crescent towards $b$ and two phase spaces can be defined within the available data of figure 3 as the smallest distances and frequencies of event whole iterations and interactions (phase I – figure 7), and depending on the empirical observation over position $S_n$, which increases to the time length of iterations and interactions of the event as a non-canonical phase space, the maximum exponential growth distance can be identified as the $b$ constant (phase II – figure 7).

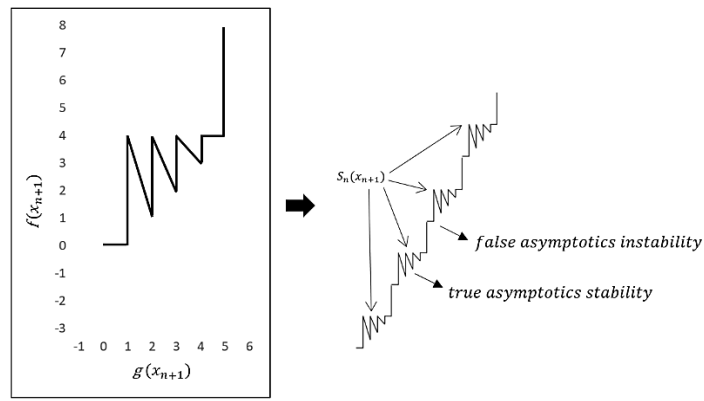

**Fig. 7.** Nonlinear time series whole representation of $\sum_{S_n+1}^{5} f(g(x))$ for a theoretical interval, identifying two defined phase spaces occurring, divided by phases I and II evidencing the distribution of the iterated phenomena between distinct functions $f(x)$ and $g(x)$ for a dynamics of true asymptotic stability and false asymptotic instability. Overall the event presents an exponential Lyapunov equilibrium, continuously iterating with the unobservable pattern formation of the stability property (figure 8).

Analyzing phase spaces in the iterated functions at a dynamic fixed point, the vectors variations that can be obtained by a fixed point dynamics in terms of time/space/other physical property observations. Representing in figure 8, a sequence of dynamic fixed points that occur as in figure 3 is presented, and it is possible to observe the iterated functions convergence illustration.

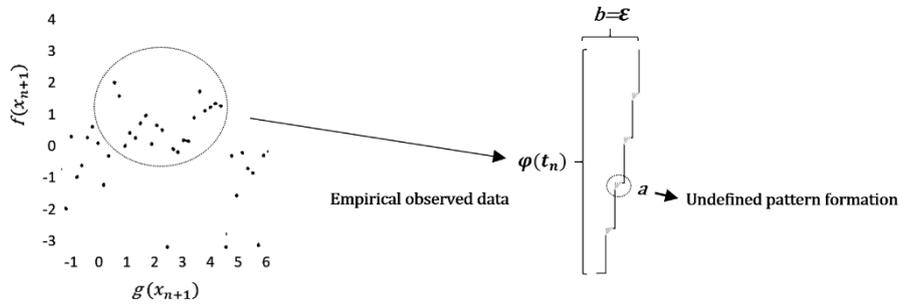

**Fig. 8.** Illustrating figure 2 and 3 information: false asymptotic instability behavior at iterating functions with Lyapunov $a$ stability and $b$ exponential stability in nonlinear time series.

By observing figure 3 properties, empirical object of this research, it is possible to realize that the dynamics between asymptotic true and false events remains as a defined phase space at figure 9.

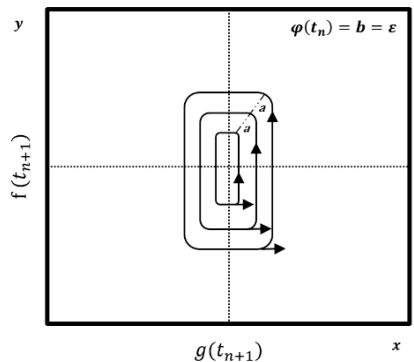

**Fig. 9.** Phase portrait of the example shown in this research showing possibility of Lyapunov exponential stability formation for some empirical phenomena.

It is not unusual asymptotic analysis are related to time aspects, leading to conclusions that confirm the asymptotic instability behavior rather than time effects over it. Empirical data observations can easily be interpreted as an asymptotic instability expression of the phenomena while time can be the true cause of phenomena expression of stability [4], making asymptotic analysis and conclusions rather a false observation of the experiment, simulation or natural observation.

# 4   Discussion

Globally speaking, the entire event presents a Lyapunov exponential stability due to the constant *b* and average orbits interaction growth towards the nominal value of constant *b*, defined by specific dynamical fixed points.

In several events involving iterations, such as physics, biology and chemistry phenomena and related knowledge, iteration events are manifestations of repetition in the order or disorder of elements that constitute an iteration [30,31]. Although the cause of iterations as well as their behavior are key topics as the coupling functions [32] in the search for iterated events. An important aspect raised in this research reflects on the discretization of iterated events in terms of knowing how they behave and how to reproduce the same event under new experimental conditions. Empirically, many events are difficult to predict [4], which also affects the possibility of identifying a possible convergence in the proposed solutions of stability [33] as it can be seen in the empirical results and methods of Bastin *et al* [34], Bingtuan *et al* [35], Di Francesco *et al* [36] and Maron *et al* [37].

One important aspect not discussed at the methodology and results sections is that even if the system has influence of new variable's input, thus modifying system stability behavior, that was described in this research as a possible false instability effect, the global behavior through time assumes possibly a new pattern formation, variances that change time to time, thus characterizing the stability aspect defined in this research for a linear or nonlinear time series. Similar approaches were also investigated to cover nonlinear exponential function behavior rate through time, as defined in Chadwick *et al* [38], but not defining time as an empirical constraints and database for the mathematical analysis of the event. Also, at Zhou [39], auto distance correlation function was defined from Szekely et al. (2007), to measure nonlinear time series, however, not defining coupling functions and its physical topological properties in a broader framework of analysis.

If there is an initial definition of a maximum event with probabilities, the expected value of the frequency with which these probabilities occur will be limited to a maximum and a minimum of variation of $f(g(S_n))$, thus defining itself by the functions convergence to the given fixed point and its orbits proximity, being it an artificial or natural expression. This same approach, not considering probabilities measure, was performed by Dionisio *et al* [40] for financial market (stock indexes) nonlinear time series, however, even with empirical constraints considered, and also notes on the time lengths limitations to the observation of the event, dynamic fixed points are considered as a dependent rate and expression of variables involved, hence defining the nonlinear phenomenon as stochastic in its finite analysis and predictable within a long term analysis (time lengths). However, the model considers the concept of dynamic fixed points without the evaluation and approaches of how the mechanisms in which coupling functions might assume in the given event, can be tracked or even manipulated. This might lead to the assumption that dependent variables are time-varying expressions in which the possibility of prediction is directly proportional to the use of a surrogate data, empirical values and autonomous observations, without the induction of dynamic fixed points existence and stability pattern formation.

These limits define that no matter how much empirically the number of elements in the process increases or assume an ordered arrangement, the maximum relative frequency of convergence does not exceed its limits since the initial function is already defined at a fixed point, that is, of the rate of convergence. That aspect of engineering solutions to complex systems has great importance for contemporary science in many fields of knowledge. This phenomenon can be very related to weaker and strong law of large numbers properties where initial conditions of phenomenon can present constant instability expression due to short time lengths in observation, that reflects as well the number of observations in a given phenomenon. As pointed at methodology section, analogously, the weak and strong law of large numbers if considered only theoretically, it does not express the same empirical aspect of convergence that does necessarily imply time dependency to prove overall exponential or not exponential stability of a given phenomenon studied.

For all interactions that may not have an initial solution given and have a high rate of instability, the empirical aspect can express the tendency of the event to balance oscillations as it does not reach infinite randomness or present great difficulty to predict. But in the other hand, events with low rate of convergence, affect the definition of the oscillation convergence more, because time allows the expression of variations of possible outcomes or unsolvable solutions that intensify through time. This asymptotic instability effect can't be observed theoretically with traditional mathematical definitions since theoretical procedures can't be time accelerated with empirical domain constraints and excluding this alternate dimension of experiment, it leads to a prove by simple deduction that the phenomena property of false asymptotic instability can't exist.

In the research done of Thomas et al [41], experimentally, was proved to the contrary of the dynamic fixed points concept. However, with new empirical expressions of these same events, the asymptotic instability effect might be observable within the noises encountered among empirical and theoretical definitions with solutions not yet proposed for this apparent instability detected due to empirical constraints and the type of experimentations of the study and this should be treated as far as new empirical experimentations, simulations or frameworks can be understood, within a new form of analysis.

If we consider, for example, atmospheric events, in a temperature range between distinct air masses, to be considered as a set A cold mass and a set B hot mass, the interaction between these events necessarily generates a region of instability C, caused mostly by iteration properties. This example helps to illustrate that the dynamics of continuously iterating atmospheric particles, according to the methodological definitions of this research, will express behavior in which regions of the physical space under which, it creates a dynamic fixed point, express symmetric sets A and B, in turn modifying themselves, generating densities of iterations and interactions over space, formed to the prior evolution of the event, by a growth or decrease move of all dynamical fixed points over the time. Dynamical fixed point reveals that regions of space in which particles will have the highest and lowest time length of interaction and iteration to achieve exponential stability, will affect the physical expression effect of particles in physicochemical aspects. The duration of the time length of iterations and interactions in different regions of space allows both sets A and B (hot and cold air mass) to express higher or lower time influence on the event with expressions of greater or lesser influence on the linear and nonlinear dynamics of the physical and chemical phenomena effects of the event.

One last important aspect of this research resembles on the $x$ nominal value considered for analysis of convergence. In respect to the plural form of the physical world, not only the time, as used in this research, could resemble connectivity and stability property of a given phenomenon. It means the physical space as well as other physical properties or nominal values of nonphysical parameters can be used as a tool to identify the same research aspect of convergence. One way of empirically observing the expression of iterated functions in connectivity on time, similar to the descriptions presented in this research, would be in figure 8. In the left image [42,43] a chemical event of wheat is represented in which the time length of iterations and interactions has a false asymptotic instability at the beginning of the event A (higher time length of iterations that gives a certain physicochemical property), and asymptotic stability after the initial phase of event B (the loss of physicochemical properties that give rigidity to the material, which is understood to be an asymptotic stability of the event). And in the image at bottom, the closed pine cone biological structure in its fractals and shape [44] can be observed for its asymptotic stability (regular time lengths) at the center of the structure, obtaining maximum interactions and determined time length of iterations, and in the final portion of the structure (tapered) it is possible to observe the false asymptotic instability (higher expression of a given function on time) in which the interactions between the functions become smaller and, consequently, the frequencies on time with which each iterated function expresses become larger.

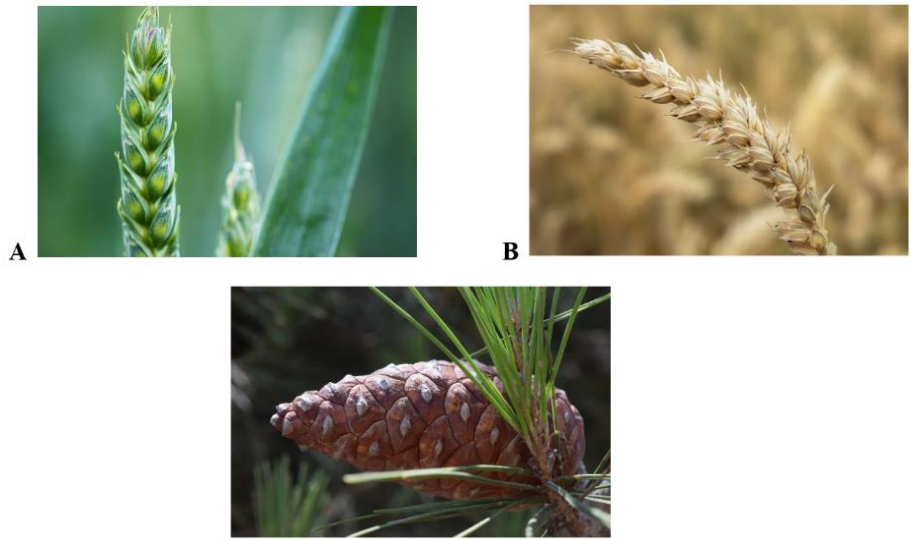

**Fig. 8.** Representation of an event with a possible Lyapunov exponential stability and its stable and unstable regions of convergence.

Note that there might be some variations on the fractal and the form of the closed pine cone as well as rigidity or bendiness of wheat structure due chemical or biological factors such as defective proteins, genes or other environmental conditions. But it does not change the mathematical modeling of describing in this research.

## 5    Conclusion

One first conclusion of the theoretical research conducted, was to demonstrate that in nonlinear time series analysis the asymptotic instability observed with empirical data sets, if time accelerated, this feature reveals itself as an asymptotic stability behavior oriented by an attractive distribution. One technical conclusion arising from it and the results section is about when investigating the empirical behavior of complex events, many conducted experiments can also fail when trying to identify asymptotic stability behavior on events, mainly defining this asymptotic stability behavior as unstable, unpredictable and without any prior pattern formation.

Also, deriving from this, the mathematical observation through theoretical point of view can fail in identifying the false instability effect due methodology constraint caused by the lack of inductive inference. This point leads the methodology proposed in this article as an observation to most scientific data concerning fixed point stability that is disconnected of empirical data sets.

A second conclusion states that considering that iterated complex variables can possibly not present a pattern formation for short time lengths, the average distance between fixed points is also followed by a low rate of convergence (instability), that in terms of scientific empirical interpretation can be understood as the phenomenon not presenting a high level of interactions expected to be in convergence to any given fixed point, but oriented to the maximum and minimum range of metric spaces defined within it (stability). Though, these interactions are existent in the extent of time of empirical observation be enough greater than the periodic time in which complex variables interact itself resulting in many infinite oscillation expression patterns, also leading to the concept of dynamical fixed point existence. This could lead to the statement that if a phenomenon presents a very high level of complexity, the outcomes of convergence also present higher time periods to be able to expand it into a new degree of pattern formation. This time-varying patterns of convergence and non-convergence at specific time lengths can also be confirmed theoretically and empirically by many researches performed by Aneta Stefanovska and PVE McClintock, as well as the axioms of the Shannon's theory of communication, the law of great numbers physical properties and monotone functions as well.

Also another aspect of detecting convergence or non-convergence, distinct manifolds of a given phenomenon can present constrained expression over time lengths, leading the observer to the use of other physical/chemical/biological exponential oriented constant (nonautonomous differential equations).

**Funding:** This research received no external funding.
**Conflicts of Interest:** The author declares no conflict of interest.